\documentclass[letter,bibyear]{aa} 

\usepackage{txfonts}
\usepackage{epsfig}
\usepackage{amsmath}
\usepackage{natbib}
\usepackage{multirow}
\usepackage{color}
\usepackage{ulem}
\usepackage{longtable}
\usepackage{xltabular}
\usepackage{graphicx}
\usepackage{epstopdf}
\usepackage{booktabs}
\usepackage{txfonts}
\usepackage{soul}

\newcommand{\jms}{J.~Mol.~Spectrosc.}

\newcommand{\doce}{10$^{12}$\,cm$^{-2}$}

\begin{document}

\title{Discovery of 1$H$-cyclopent[$cd$]indene ($c$-C$_{11}$H$_8$) in TMC-1 with the QUIJOTE line survey: A new three-ringed polycyclic aromatic hydrocarbon\thanks{Based on 
observations carried out
with the Yebes 40m telescope (projects 19A003,
20A014, 20D023, 21A011, 21D005, and 23A024). The 40m
radio telescope at Yebes Observatory is operated by the Spanish Geographic Institute
(IGN, Ministerio de Transportes y Movilidad Sostenible).}}

\author{
R.~Fuentetaja\inst{1},
C.~Cabezas\inst{1},
M.~Ag\'undez\inst{1},
B.~Tercero\inst{2,3},
N.~Marcelino\inst{2,3},
P.~de~Vicente\inst{2},
J.~Cernicharo\inst{1}
}

\institute{Dept. de Astrof\'isica Molecular, Instituto de F\'isica Fundamental (IFF-CSIC),
C/ Serrano 121, 28006 Madrid, Spain. \newline \email r.fuentetaja@csic.es, jose.cernicharo@csic.es
\and Observatorio de Yebes (IGN), Cerro de la Palera s/n, 19141 Yebes, Guadalajara, Spain.
\and Observatorio Astron\'omico Nacional (OAN, IGN), C/ Alfonso XII, 3, 28014, Madrid, Spain.
}

\date{Received; accepted}

\abstract{
We report the detection of the polycyclic aromatic hydrocarbon (PAH) 1$H$-cyclopent[$cd$]indene ($c$-C$_{11}$H$_8$) in TMC-1 with the QUIJOTE line survey. We detected 22 independent lines corresponding to 88 rotational transitions with quantum numbers ranging from $J$=19 up to $J$=24 and $K_a \le 5$ in the Q-band range. The identification of this new PAH was based on the agreement between the rotational parameters derived from the analysis of the lines and those obtained by quantum chemical calculations. The column density derived for 1$H$-cyclopent[$cd$]indene is (6.0\,$\pm$\,0.5)\,$\times$\,\doce, with a rotational temperature of 9 K. Its abundance is high, as is that of the rest of the PAHs, but it is the lowest of all those detected to date in TMC-1, being 2.66 times less abundant than indene and 4.66 times less than phenalene. This result will help us to better understand the growth of five- and six-membered rings in dark clouds. Chemical models explaining their formation through the bottom-up model are still very incomplete and require further experimental and theoretical effort. Even so, the most likely formation reactions would occur between the smallest rings with small hydrocarbons; the most probable reaction for the formation of cyclopentindene is that between indene and C$_2$H, C$_2$H$_3$, and/or their cation.
}

\titlerunning{1$H$-cyclopent[$cd$]indene in TMC-1}
\authorrunning{Fuentetaja et al.}

\maketitle

\section{Introduction}

\begin{figure*}
    \centering
    \includegraphics[width=1\textwidth]{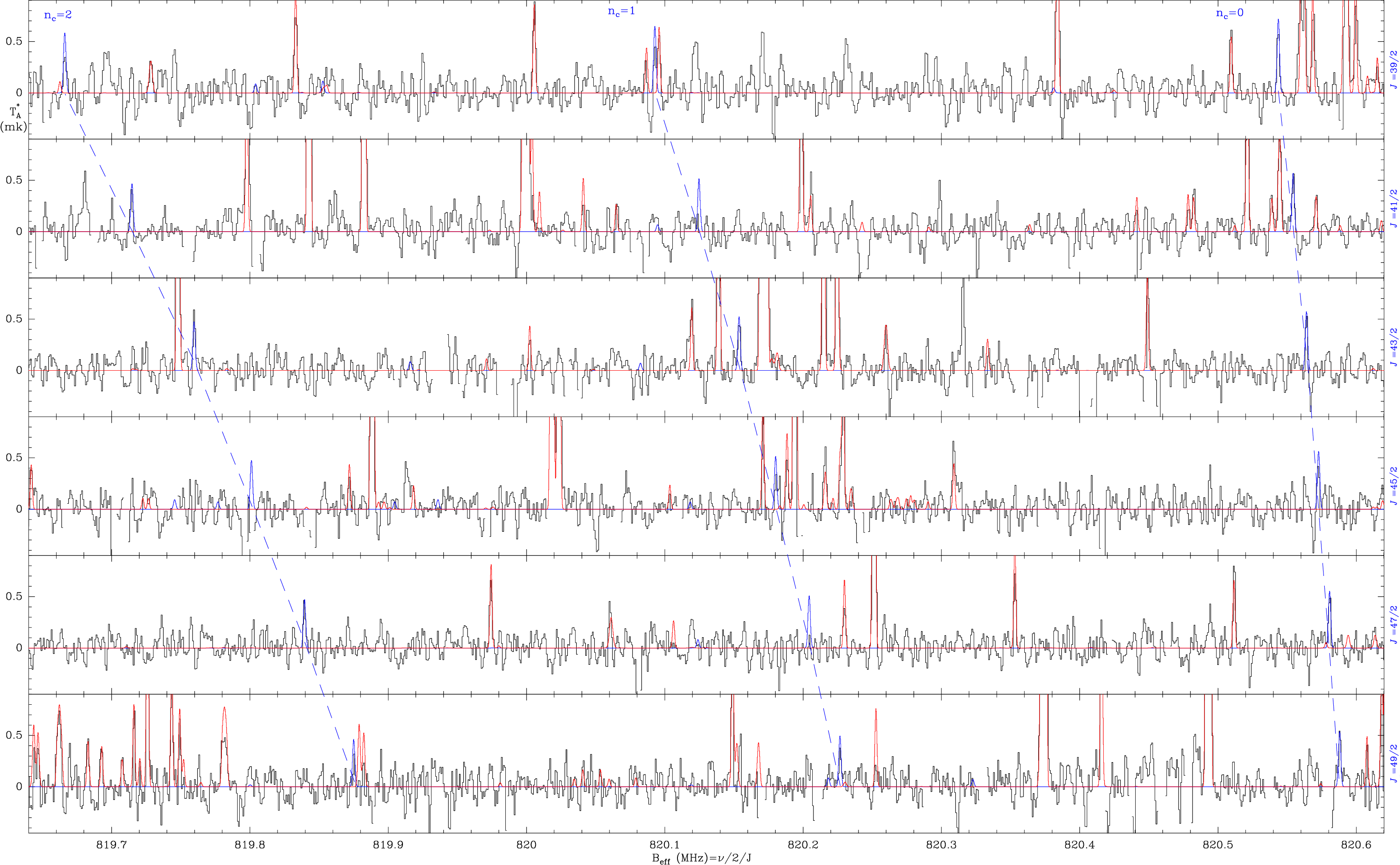}
    \caption{Modified Loomis--Wood diagram of some of the observed lines of cyclopentindene. The abscissa corresponds to the value of the rotational constant, which has been fixed in this plot to values between 819.6 and 820.6 MHz. The ordinate is the antenna temperature, corrected for atmospheric and telescope losses, in millikelvin. The frequency represented in each box corresponds to 2 B$_{rot}$~$J$, where $J$ = $J_u$ + 1/2. The red lines represent a synthetic spectrum of the lines detected with the QUIJOTE data. The blue solid lines correspond to the new detected lines in this work, while dotted lines correspond to transitions with the same $K_a$ (for all transitions involved, see Table \ref{tab:lineparameter}).}
    \label{fig:LW}
\end{figure*}

The dark cloud TMC-1 has become the best astrophysical source for studying the chemistry of the interstellar medium; it is a natural laboratory where we can evaluate our chemical models. In recent years, more than 70 new molecules have been identified in TMC-1, significantly expanding the known inventory of chemical complexity.

The sensitivity of the Q-band Ultrasensitive Inspection Journey to the Obscure TMC-1 Environment (QUIJOTE) survey has been decisive for these results. It has allowed the detection of species with very low dipole moments \citep{Cernicharo2021a, Cernicharo2024a}, whose rotational lines have weak intensities, and therefore require long integration times, as in the case of long carbon chains. Among the latest discoveries made with QUIJOTE, we note  NC$_3$S and HC$_3$S \citep{Cernicharo2024b}; HNC$_5$ \citep{Fuentetaja2024}; HC$_3$N$^+$, HC$_5$N$^+$, and HC$_7$N$^+$ \citep{Cabezas2024, Cernicharo2024c}; CH$_2$CHCHS \citep{Cabezas2025a}; CH$_3$CHS \citep{Agundez2025}; and HCCCHCN \citep{Cabezas2025b}.

Among the species detected, polycyclic aromatic hydrocarbons (PAHs) are of particular interest. They are one of the most common types of organic molecules in the interstellar medium and were the first plausible detection using aromatic infrared bands in the mid-infrared range \citep{Allamandola1985}. However, this technique has not allowed any specific species to be identified. The first unequivocal detection of a PAH occurred in dark clouds and were the cyano-naphthalene derivatives, reported by \cite{McGuire2021}, which established an observational precedent for the presence of polycyclic aromatic species in \mbox{TMC-1}. Subsequently, a considerable number of cyano functionalized PAHs were detected in the same cloud, including 2-cyanoindene, cyano-acenaphtylenes, cyano-pyrenes, and cyano-coronene \citep{Sita2022,Cernicharo2024a,Wenzel2024,Wenzel2025a,Wenzel2025b}. The inferred abundances of the corresponding non-functionalized forms were rather high, in the range 10$^{-9}$-10$^{-8}$ relative to H$_2$. A clearer picture emerged with the direct detection of a few non-functionalized PAHs, such as indene \citep{Cernicharo2021a,Burkhardt2021} and phenalene \citep{Cabezas2025c}. These molecules could be detected in spite of their low dipole moments, well below those of the cyano derivatives observed in \mbox{TMC-1}. The abundances determined for indene and phenalene, above 10$^{-9}$ relative to H$_2$, were consistent with those estimated from the cyano functionalized forms. The observed abundances of PAHs are surprisingly high compared to previous expectations based on extrapolations or analogies with linear species. This  forces us to reconsider the role of PAHs in the TMC-1 reaction network \citep{Wakelam2008}, both in the gas phase and potentially in interaction with grains, and to improve current chemical models in order to explain their presence in the source. Furthermore, given the existence of such a large PAH as cyanocoronene (with a structure of 24 carbon atoms), we would expect to find a large number of PAHs with comparable sizes having column densities similar to those of the PAHs already detected.

Here we present the detection of 1$H$-cyclopent[$cd$]indene (hereafter $c$-C$_{11}$H$_8$; cyclopentindene), which is the first PAH with two five-atom cycles in its structure and the third unsubstituted PAH detected in the ISM. This discovery will contribute to a better understanding of the formation routes of PAHs containing five- and six-membered rings, which until now have only been studied for cyclopentadiene and indene.

\section{Observations}
The observational data used belong to QUIJOTE$^1$ \citep{Cernicharo2021b, Cernicharo2024a}, a Q-band spectral line survey of TMC-1 performed with the Yebes 40 m telescope at coordinates $\alpha_{J2000}=4^{\rm h} 41^{\rm  m} 41.9^{\rm s}$ and $\delta_{J2000}=+25^\circ 41' 27.0''$, corresponding to the cyanopolyyne peak (CP) in TMC-1. The receiver was developed within the Nanocosmos project,\footnote{\texttt{https://nanocosmos.iff.csic.es/}} consisting of two cooled high electron mobility transistor (HEMT) amplifiers covering the 31.0--50.3 GHz band with horizontal and vertical polarization. Fast Fourier transform spectrometers (FFTSs) with 8 × 2.5 GHz with a spectral resolution of 38.15 kHz provide the whole coverage of the Q-band in both polarizations. A complete description of the system is provided by \citet{Tercero2021}. The QUIJOTE survey data presented here were obtained over multiple series of observation beginning in November 2019. All observations were performed in the frequency-switching observing mode with a frequency throw of either 10 or 8 MHz.
The total observing time on source for data taken with frequency throws of 10 MHz and 8 MHz was 772.6 and 736.6 hours, respectively. Hence, the total observing time of the QUIJOTE line survey was 1509.2 hours. Actually, the achieved sensitivity ranged from 0.06 mK at 32 GHz to 0.18 mK at 49.5 GHz, about 100 times better than the earlier Q-band TMC-1 line surveys \citep{Kaifu2004}. A detailed account of the QUIJOTE line survey and the data analysis is given by \citet{Cernicharo2021b, Cernicharo2022}. The main-beam efficiency measured during our 2022 observations varied from 0.66 at 32.4 GHz to 0.50 at 48.4 GHz \citep{Tercero2021} and across the Q band follows $B_{\rm eff}=0.797,\exp[-(\nu(\mathrm{GHz})/71.1)^2]$. The forward efficiency of the telescope is 0.97. The half-power beam size is 54.4$''$ at 32.4 GHz and 36.4$''$ at 48.4 GHz.

\section{Results} \label{results}

\begin{figure}
    \centering
    \includegraphics[width=0.8\linewidth]{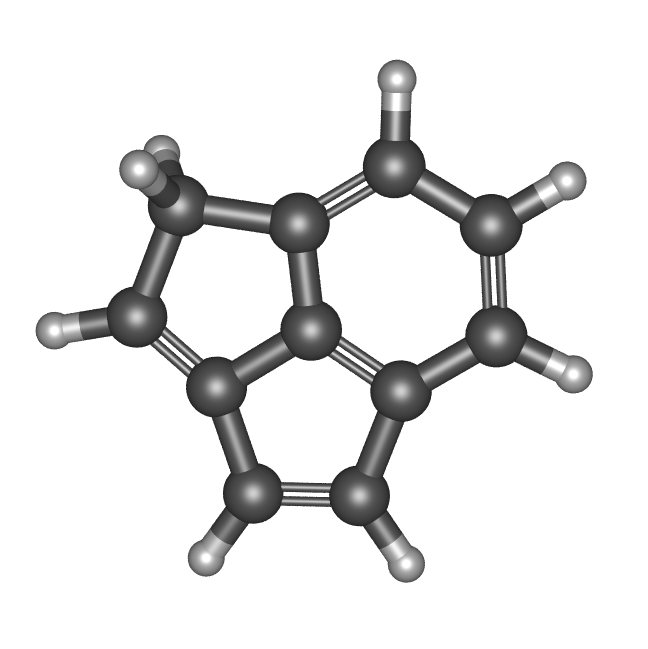}
    \caption{Optimized geometry of cyclopentindene.}
    \label{fig:geometry}
\end{figure}

Similar to the methodology employed for the detection of other PAHs with the QUIJOTE line survey \citep{Cernicharo2024a}, the analysis began with the identification of a harmonic series from the spectral data, carried out with a modified Loomis–Wood diagram that uses semi-integer $J+1/2$ values (see Fig. \ref{fig:LW}). The frequency plotted in each box corresponds to 2C($J+1/2$), thus representing the $C$ rotational constant on the abscissa axis. These series can be explained by high $J$ values, such as those observed in QUIJOTE, using the equations developed by \citet{Watson2007}, which provide an equation for the energy in asymptotic cases for asymmetric molecules within the rigid rotor approximation.

We found a new spectral patter centred around 820.5 MHz, which corresponds to transitions $J$=$K_c$ (with $K_a$=0 and 1) collapsed at the same frequency (see Table \ref{tab:lineparameter}). From there, with the help of the SPFIT programme \citep{Pickett1991}, and using a Watson's Hamiltonian with the A-reduction and the $III^l$ representation \citep{Watson1977}, we were able to fit all the observed transitions to higher $K_a$ (see Fig. \ref{fig:LW}). The derived spectroscopic constants reproduce the observed series with a rms = 6.2 kHz.

Once we have the rotational transitions fitted, we can obtain the molecular constants to identify the molecular carrier responsible of these lines. First of all, the Ray's parameter value ($\kappa = (2B-A-C)/(A-C)$) tells us what type of asymmetric rotor the carrier is. The value obtained for $\kappa$ is 0.36, which indicates that the molecule is slightly oblate. PAH derivatives, such as the cyano derivatives detected from pyrene, naphthalene, or acenaphthylene, have a prolate structure, so in principle we  only consider pure PAHs. Furthermore, the inertia defect ($\Delta_c = I_C -I_B -I_A$) provides the planarity of the molecule. A value of 0 or close to it would mean that the molecule is almost planar. We obtained a value $\sim$3.3351 amu $\mathring{A}^2$, which is very similar to that obtained in the case of phenalene or indene. This indicates that the molecule is mainly planar with two hydrogens outside the plane \citep{Gordy1984}. Finally, we can use the values of the obtained rotational constants to compare them with other PAHs detected in TMC-1, which  indicates the size of the molecule. 

Indene has the closest values, but higher than the ones obtained here, which indicates that the new molecule is larger than indene. On the other hand, we see that phenalene has lower values, so we would expect its size to be less than three cycles of six carbon atoms. Within the family of molecules with two six-atom cycles and one five-atom cycle, fluorene does not have a slightly oblate structure; acenaphthylene does not have any hydrogen atoms outside the plane; and both, together with acenaphthene, have constants lower than those obtained. Therefore, the two main candidates we have are 1H-cyclopent[cd]indene and its 2H-cyclopent[cd]indene isomer, which has an energy 94.69129 kJ/mol higher since they are slightly oblate molecules with two hydrogens outside the plane that have two cycles of five atoms and another cycle of six atoms.

We performed quantum chemical calculations and considered all the information mentioned above. The geometries of the PAHs were optimized using the B3LYP/6-311++G(d,p) level of theory \citep{Becke1993,Frisch1984}, and the Gaussian 16 package \citep{Frisch2016}, which has been shown to give good results for this type of molecule \citep{Cernicharo2024a}. 
The results obtained for this molecule, both for the rotational and distortion constants and the dipole moment components, are shown in Table \ref{tab:constants}. The values of our search suggest that the best candidate is 1H-cyclopent[cd]indene, whose equilibrium geometry is shown in Fig. \ref{fig:geometry}. The very good agreement between these values and those obtained experimentally confirms that the harmonic series found corresponds to this molecule.

Initially, the observed lines were adjusted taking into account only a-type transitions, but the molecule presents similar values in the $\mu_a$ and $\mu_b$ components of its dipole moment, 0.6 and 0.8, respectively. Therefore, there is a significant contribution from the b-type transitions that were included in the adjustment. For each detected line, we have four transitions collapsed at the same frequency, two of a-type and two of b-type, with $K_c$=$J$-$K_a$ and $K_c$=$J$-$K_a$-1.

\section{Discussion}

Line identification in this work was performed using the MADEX code \citep{Cernicharo2012}. The intensity scale used in this study is the antenna temperature (T$_A^{\ast}$). Consequently, the telescope parameters and source properties were used when modelling the emission of the different species to produce synthetic spectra on this temperature scale. For this work we assumed a velocity for the source relative to the local standard of rest of 5.83 km s$^{-1}$ \citep{Cernicharo2020}. The source was assumed to be circular with a uniform brightness temperature and a radius of 40'' \citep{Fosse2001}.

\begin{figure}
    \centering
    \includegraphics[width=1\linewidth]{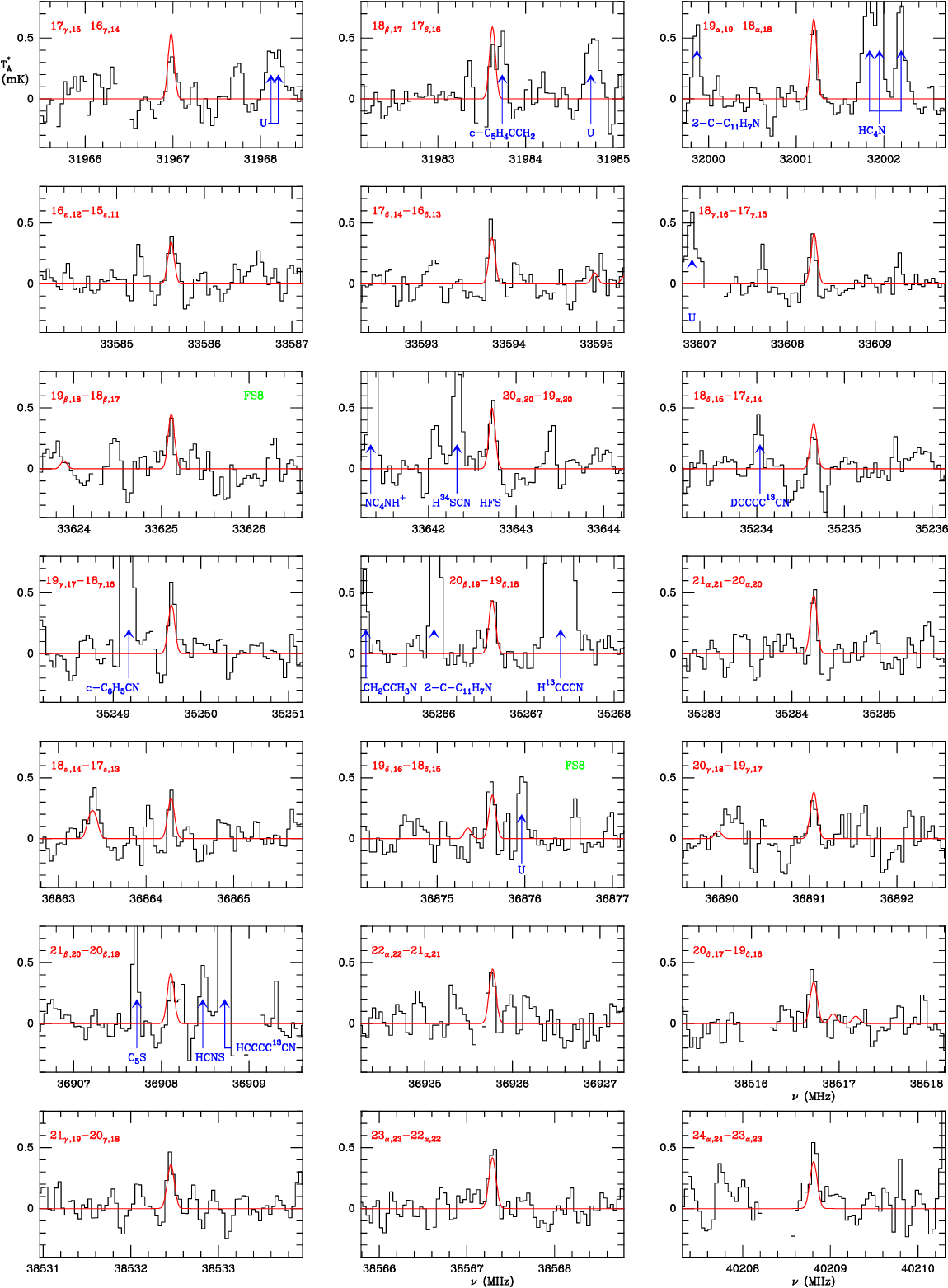}
    \caption{Some of the cyclopentindene lines observed in TMC-1 with QUIJOTE. The black line represents the survey data, while the red line represents the synthetic spectrum calculated for a column density of 6$\times$\doce. The green labels indicate the transitions for which only  the 8 MHz frequency-switching data were used. The blanked channels correspond to negative features resulting in the folding of the frequency-switched spectra. Each box shows four collapsed transitions corresponding to each line, where $\alpha$, $\beta$, $\gamma$, $\delta$, and $\epsilon$ are $K_a$=0,1, $K_a$=1,2, $K_a$=2,3, $K_a$=3,4, and $K_a$=4,5, respectively (see Table \ref{tab:lineparameter} for more details about transitions). The unidentified lines or other molecules already detected in the source appear in blue. The abscissa corresponds to the rest frequency assuming a local standard of rest velocity of 5.83 km s$^{-1}$. The ordinate is the antenna temperature in millikelvin.}
    \label{fig:lines}
\end{figure}

To obtain the column density, we assumed that cyclopentindene is close to thermalization (T$_{rot}$ = 9 K), based on benzonitrile and the other PAHs detected at the source \citep{Cernicharo2023, Cernicharo2024a}. We derived a column density of (6.0\,$\pm$\,0.5)\,$\times$\,\doce (see Fig. \ref{fig:lines}). This value is the lowest detected in pure PAH in TMC-1. The abundance of cyclopentindene is 2.66 and 4.66 times lower than those of indene and phenalene, respectively. These values were obtained at a T$_{rot}$ of 10 K for indene and 7.9$\pm$1.2 K for phenalene \citep{Cernicharo2021a,Cabezas2025c}.

The processes involved in the formation of PAHs are unclear. Chemical models are being improved to try to explain the observed abundance. There are two main hypotheses regarding their formation. The first is the bottom-up model, which proposes that PAHs form in cold dark clouds from smaller hydrocarbons. These molecules would grow and form increasingly larger cycles. The second hypothesis is  the top-down model, which considers that small PAHs originate from the fragmentation of very large PAHs coming from outside the cloud. In TMC-1, observational data favour the bottom-up model. Spatial distributions, for example that of benzonitrile, coincide with those of cyanopolyynes and other reactive species, and therefore the data point to local formation from carbon chains, radicals, and cations detected at the source \citep{Cernicharo2023}.

The chemical pathways involved probably combine neutral-neutral and ion-neutral reactions, although the exact routes are not yet clear. For example, for the formation of single-cycle molecules, the reaction between l-C$_3$H$_3^+$ and C$_2$H$_4$ has been recently shown to be rapid although it does not produce cyclic C$_5$H$_7^+$, a direct precursor of cyclopentadiene, but another cyclic cation, such as C$_5$H$_5^+$ \citep{Mallo2025}.

For three-ringed species, specific bottom-up routes have been proposed. \citet{Kaiser2021} proposed the addition of five-carbon cycles through the reaction of vinyl derivatives of PAHs with CH. In acenaphthylene, direct reactions of C$_2$H or H$_2$CCH (not detected yet) with naphthalene or C$_3$H with indene are suggested as potential formers, but further studies are needed to assess the viability of these routes \citep{Cernicharo2024a}. In the case of the pure PAH phenalene, a viable ion neutral pathway has been identified based on radiative association C$_{12}$H$_8$ + CH$_3^+$ $\rightarrow$ C$_{13}$H$_{11}^+$ + h$\nu$, followed by dissociative recombination to yield C$_{13}$H$_{10}$. This pathway is evaluated as exothermic and without a barrier in the association step, highlighting the possible expanded role of ion-molecule chemistry in PAH formation \citep{Cernicharo2022, Cabezas2025c}. This latter neutral ion pathway has been proposed for all cyclic species with five and six atoms and for the PAHs detected so far in TMC-1 \citep{Loru2023,Cernicharo2021a, Cernicharo2021c, Cernicharo2021d, Cernicharo2022, Cernicharo2024a}.
Analogous to these PAHs, the reaction of C$_2$H (or its cation) with indene could be a viable formation pathway for cyclopentindene. Despite all this information, theoretical or experimental processes are still needed to study the possible formation pathways in order to correctly describe their presence.

\section{Conclusions}
In this work we reported the first detection in the ISM of cyclopentindene, a new three-ringed molecule, based on the visual search of a harmonic pattern and quantum chemistry calculations. The column density derived is 6\,$\times$\,\doce. Its abundance is the lowest of all pure PAHs detected in TMC-1 when compared to indene and phenalene. Possible formation pathways for PAHs were discussed, but the chemical models are currently incomplete and further experimental and theoretical studies are needed to properly understand the formation processes and abundances of these species. Finally, the detection of cyanocoronene suggests that in the future we may be able to detect more intermediate-sized PAHs that will aid our understanding of the role these molecules play in the chemistry of the ISM.

\begin{acknowledgements}

The present study was supported by Ministerio de Ciencia e Innovación of Spain (MICIU) for funding support through projects PID2023-147545NB-I00 and PID2022-137980NB-100. Also thank ERC for funding through grant ERC-2013-Syg-610256- 312 NANOCOSMOS. 

\end{acknowledgements}

\normalsize

\begin{appendix}
\onecolumn
\section{Rotational constants}

\begin{table}[h]
    \centering
    \caption{Experimental and theoretical molecular constants of cyclopentindene.}
    \begin{tabular}{|c|c|c|c|}
         \hline
         Parameter & 1H-cyclopent[cd]indene$^a$ & 2H-cyclopent[cd]indene$^a$ &TMC-1 ($III^l$) \\
         \hline
$A$/MHz           &  1809.3791 & 1769.6150  & 1804.613(71)$^b$    \\
$B$/MHz           &  1490.4924 & 1510.4163   & 1490.725(56)    \\
$C$/MHz           &  821.38945 & 818.97200   & 820.77500(68)  \\
$\Delta_J$/Hz     &  45.05    &  46.95    & 51.0(45)     \\
$\Delta_{JK}$/Hz  & -59.98    &  -66.53   & -66.5(50)     \\
$\Delta_K$/Hz     &   19.75   &  24.42   &  [19.75]$^c$         \\
$\delta_J$/Hz     &  -3.72    &  -4.73   &  [-3.72]         \\
$\delta_K$/Hz     &  24.74    &   20.26 &  [24.74]        \\
         \hline
         $N_{trans}^d$,$N_{lines}^e$ &  &   & 88, 22 \\
         $J_{max}, K_{a,max}$  & &   & 24, 6\\
         $\sigma (kHz)^f$,$\sigma_w^g$ &  &   & 6.2, 0.65 \\
         $\mu_a$,$\mu_b$ &   0.60, 0.80  &  0.28, 1.12 & \\
         \hline
    \end{tabular}
    \label{tab:constants}
    \tablefoot{\\
    \tablefoottext{a}{B3LYP/6-311++G(d,p) level of theory.}\\
    \tablefoottext{b}{The uncertainties (in parentheses) are in units of the last significant digits.} \\
    \tablefoottext{c}{Values in square brackets have been kept fixed to the theoretical values.}\\
    \tablefoottext{d}{Total number of rotational transitions.}\\
    \tablefoottext{e}{Total number of independent frequencies.}\\
    \tablefoottext{f}{Standard root mean square deviation of the fit in kHz.} \\
    \tablefoottext{g}{Weighted root mean square deviation of the fit, unitless.} \\
    }
\end{table}

\section{Derived line parameters}
The line parameters derived for this work were obtained by fitting a Gaussian line profile to the observed data, using the software Class (GILDAS package). We used a window of $\pm$15 km s$^{-1}$ around the V$_{LSR}$ (5.83km s$^{-1}$) of the source for each transition. Negative features in the folding of the frequency switching data were blanked before baseline removal. Several of these transitions can be observed in Fig. \ref{fig:lines}.

\tiny
\begin{xltabular}{\linewidth}{@{} c c c c c c @{}}
    \caption{Observed line parameters for cyclopentindene.}
    \label{tab:lineparameter}
    \\ \hline
    Transition & $\nu _{obs}$$^a$ & $\shortmid$$\nu _{obs-calc}$$\shortmid$  & $\int$ $T_A^*$ dv $^b$ & $\Delta$v$^c$ &   T$_A^*$$^d$  \\
     &  MHz         &  MHz  & mK km s$^{-1}$&km s$^{-1}$  &   mK     \\
    \hline
    \endfirsthead
    \caption[]{continued.}\\
    \hline
    Transition & $\nu _{obs}$$^a$ & $\shortmid$$\nu _{obs-calc}$$\shortmid$  & $\int$ $T_A^*$ dv $^b$ & $\Delta$v$^c$ &   T$_A^*$$^d$  \\
     &  MHz         &  MHz  & mK km s$^{-1}$&km s$^{-1}$  &   mK     \\
    \hline
    \endhead
    \endfoot
    17$_{2,15}$-16$_{2,14}$  &  31966.982$\pm$0.010 & 0.004 &    0.29$\pm$0.09 &   0.85$\pm$0.27 & 0.32$\pm$0.10 \\
    17$_{3,15}$-16$_{2,14}$ & & & & &\\
    17$_{2,15}$-16$_{3,14}$ & & & & &\\
    17$_{3,15}$-16$_{3,14}$ & & & & &\\
    \hline
    18$_{1,17}$-17$_{1,16}$ &  31983.619$\pm$0.012 & 0.001 &    0.28$\pm$0.09 &   0.55$\pm$0.21 & 0.47$\pm$0.10 \\
    18$_{2,17}$-17$_{1,16}$ & & & & &\\
    18$_{1,17}$-17$_{2,16}$ & & & & &\\
    18$_{2,17}$-17$_{2,16}$ & & & & &\\
    \hline
    19$_{0,19}$-18$_{0,18}$ &  32001.202$\pm$0.010 & $<$0.001  &  0.52$\pm$0.07 &   0.73$\pm$0.12 & 0.68$\pm$0.10 \\
    19$_{1,19}$-18$_{0,18}$ & & & & &\\
    19$_{0,19}$-18$_{1,18}$ & & & & &\\
    19$_{1,19}$-18$_{1,18}$ & & & & &\\
    \hline
    16$_{4,12}$-15$_{4,11}$ &  33585.632$\pm$0.010 & 0.003 &   0.33$\pm$0.06 &   0.87$\pm$0.17 & 0.36$\pm$0.08 \\
    16$_{5,12}$-15$_{4,11}$ & & & & &\\
    16$_{4,12}$-15$_{5,11}$ & & & & &\\
    16$_{5,12}$-15$_{5,11}$ & & & & &\\
    \hline 
    17$_{3,14}$-16$_{3,13}$ &  33593.799$\pm$  0.010 & 0.003 &   0.39$\pm$0.08 &   0.64$\pm$ 0.13 &     0.57$\pm$0.10\\
    17$_{4,14}$-16$_{3,13}$ \\
    17$_{3,14}$-16$_{4,13}$ \\
    17$_{4,14}$-16$_{4,13}$ \\
    \hline
    18$_{2,16}$-17$_{2,15}$ &  33608.290$\pm$0.010 & 0.011  &  0.36$\pm$0.07 &   0.74$\pm$0.16 & 0.45$\pm$0.08 \\
    18$_{3,16}$-17$_{2,15}$ & & & & &\\
    18$_{2,16}$-17$_{3,15}$ & & & & &\\
    18$_{3,16}$-17$_{3,15}$ & & & & &\\
    \hline
    19$_{1,18}$-18$_{1,17}$~$^e$ &  33625.111$\pm$0.010 & 0.002  &   0.32$\pm$0.08 &   0.72$\pm$0.25 & 0.41$\pm$0.12 \\
    19$_{2,18}$-18$_{1,17}$ & & & & &\\
    19$_{1,18}$-18$_{2,17}$ & & & & &\\
    19$_{2,18}$-18$_{2,17}$ & & & & &\\
    \hline
    20$_{0,20}$-19$_{0,19}$ &  33642.731$\pm$0.010 & 0.003  &  0.41$\pm$0.09 &   0.73$\pm$0.19 & 0.53$\pm$0.10 \\
    20$_{1,20}$-19$_{0,19}$ & & & & &\\
    20$_{0,20}$-19$_{1,19}$ & & & & &\\
    20$_{1,20}$-19$_{1,19}$ & & & & &\\
    \hline
    18$_{3,15}$-17$_{3,14}$ &  35234.645$\pm$0.010 & 0.001 &   0.20$\pm$0.06 &   0.61$\pm$0.20 & 0.30$\pm$0.09 \\
    18$_{4,15}$-17$_{3,14}$ & & & & &\\
    18$_{3,15}$-17$_{4,14}$ & & & & &\\
    18$_{4,15}$-17$_{4,14}$ & & & & &\\
    \hline
    19$_{2,17}$-18$_{2,16}$ &   35249.661$\pm$  0.010 &  0.010 &   0.41$\pm$0.06 &   0.63$\pm$ 0.10 &     0.60$\pm$0.10\\
    19$_{3,17}$-18$_{2,16}$ & & & & &\\
    19$_{2,17}$-18$_{3,16}$ & & & & &\\
    19$_{3,17}$-18$_{3,16}$ & & & & &\\
    \hline
    20$_{1,19}$-19$_{1,18}$ &  35266.615$\pm$0.010 & 0.006  &  0.40$\pm$0.08 &   0.76$\pm$0.18 & 0.49$\pm$0.09 \\
    20$_{2,19}$-19$_{1,18}$ & & & & &\\
    20$_{1,19}$-19$_{2,18}$ & & & & &\\
    20$_{2,19}$-19$_{2,18}$ & & & & &\\
        \hline
    21$_{0,21}$-20$_{0,20}$ &  35284.252$\pm$0.010 & $<$0.001  &  0.33$\pm$0.05 &   0.50$\pm$0.10 & 0.62$\pm$0.09 \\
    21$_{1,21}$-20$_{0,20}$ & & & & &\\
    21$_{0,21}$-20$_{1,20}$ & & & & &\\
    21$_{1,21}$-20$_{1,20}$ & & & & &\\
    \hline 
    17$_{5,12}$-16$_{5,11}$ &  36863.401$\pm$  0.012 &  0.006 &  0.48$\pm$0.11 &   1.12$\pm$ 0.32 &     0.40$\pm$0.10\\
    17$_{6,12}$-16$_{5,11}$ \\
    17$_{5,12}$-16$_{6,11}$ \\
    17$_{6,12}$-16$_{6,11}$ \\
    \hline
    18$_{4,14}$-17$_{4,13}$ &  36864.275$\pm$  0.010 & 0.004  &  0.21$\pm$0.04 &   0.33$\pm$ 0.25 &     0.59$\pm$0.10 \\
    18$_{5,14}$-17$_{4,13}$ & & & & &\\
    18$_{4,14}$-17$_{5,13}$ & & & & &\\
    18$_{5,14}$-17$_{5,13}$ & & & & &\\
    \hline
    19$_{3,16}$-18$_{3,15}$~$^e$ &  36875.615$\pm$0.010 & 0.008  &  0.39$\pm$0.08 &   0.69$\pm$0.14 & 0.53$\pm$0.10 \\
    19$_{4,16}$-18$_{3,15}$ & & & & &\\
    19$_{3,16}$-18$_{4,15}$ & & & & &\\
    19$_{4,16}$-18$_{4,15}$ & & & & &\\
    \hline
    22$_{0,22}$-21$_{0,21}$ &  36925.765 $\pm$0.010 & 0.009  &  0.27$\pm$0.08 &   0.60$\pm$0.14 & 0.43$\pm$0.11 \\
    22$_{1,22}$-21$_{0,21}$ & & & & &\\
    22$_{0,22}$-21$_{1,21}$ & & & & &\\
    22$_{1,22}$-21$_{1,21}$ & & & & &\\
    \hline
    19$_{4,15}$-18$_{4,14}$  &  38504.336$\pm$  0.017 & 0.009 &   0.50$\pm$0.13 &   1.02$\pm$ 0.30 &     0.46$\pm$0.12 \\
    19$_{5,15}$-18$_{4,14}$ & & & & &\\
    19$_{4,15}$-18$_{5,14}$ & & & & &\\
    19$_{5,15}$-18$_{5,14}$ & & & & &\\
    \hline 
    20$_{3,17}$-19$_{3,16}$ &  38516.697$\pm$  0.010 & 0.005 &   0.32$\pm$0.07 &   0.65$\pm$ 0.16 &     0.47$\pm$0.08\\
    20$_{4,17}$-19$_{3,16}$ \\
    20$_{3,17}$-19$_{4,16}$ \\
    20$_{4,17}$-19$_{4,16}$ \\
    \hline
    21$_{2,19}$-20$_{2,18}$ &  38532.453$\pm$0.010 & 0.001  &  0.28$\pm$0.05 &   0.59$\pm$0.11 & 0.46$\pm$0.08 \\
    21$_{3,19}$-20$_{2,18}$ & & & & &\\
    21$_{2,19}$-20$_{3,18}$ & & & & &\\
    21$_{3,19}$-20$_{3,18}$ & & & & &\\
    \hline
    23$_{0,23}$-22$_{0,22}$ &  38567.293$\pm$0.010 &  0.002 &  0.36$\pm$0.07 &   0.64$\pm$0.13 & 0.52$\pm$0.09 \\
    23$_{1,23}$-22$_{0,22}$ & & & & &\\
    23$_{0,23}$-22$_{1,22}$ & & & & &\\
    23$_{1,23}$-22$_{1,22}$ & & & & &\\
    \hline
    23$_{1,22}$-22$_{1,21}$ &  40191.115$\pm$  0.012 &  0.009  & 0.20$\pm$0.05 &   0.56$\pm$ 0.19 &     0.33$\pm$0.09 \\
    23$_{2,22}$-22$_{1,21}$ & & & & &\\
    23$_{1,22}$-22$_{2,21}$ & & & & &\\
    23$_{2,22}$-22$_{2,21}$ & & & & &\\
    \hline
    24$_{0,24}$-23$_{0,23}$ &  40208.816$\pm$0.023 & 0.004  &  0.30$\pm$0.19 &   0.56$\pm$0.45 & 0.51$\pm$0.14 \\
    24$_{1,24}$-23$_{0,23}$ & & & & &\\
    24$_{0,24}$-23$_{1,23}$ & & & & &\\
    24$_{1,24}$-23$_{1,23}$ & & & & &\\ 
    \hline 
\end{xltabular}
    \tablefoot{\\
    \tablefoottext{a}{Adopted rest frequency (see text).}\\
    \tablefoottext{b}{Integrated line intensity in mK\,km\,s$^{-1}$.}\\
    \tablefoottext{c}{Line width at half intensity derived by fitting a Gaussian function to
the observed line profile (in km\,s$^{-1}$).}\\
    \tablefoottext{d}{Antenna temperature in millikelvin.}\\
    \tablefoottext{e}{ We obtained the line parameters using only the 8 MHz frequency switching data.}
    }

\end{appendix}
\end{document}